\documentclass[letterpaper,aps,pre,twocolumn,showcaps,superscriptaddress,amsmath,amssymb]{revtex4}

\usepackage[dvips]{graphicx}
\usepackage{psfrag}
\usepackage{color}
\usepackage{epsfig}
\usepackage{subfigure}
\usepackage[full]{textcomp}
\expandafter\let\csname equation*\endcsname\relax
\expandafter\let\csname endequation*\endcsname\relax
\usepackage{amsmath}
\usepackage{amssymb}

\begin{document}

\title{Correlation between crystalline order and vitrification in colloidal monolayers}

\author{Elisa Tamborini}
\address{Institut Lumi\`{e}re Mati\`{e}re, Universit\'{e} Lyon 1, France}
\address{Cavendish Laboratory, University of Cambridge, U.~K.}

\author{C. Patrick Royall}
\address{H.H. Wills Physics Laboratory, University of Bristol,  U.~K.}
\address{School of Chemistry,  University of Bristol, U.~K.}
\address{Centre for Nanoscience and Quantum Information, University of Bristol, U.~K.}

\author{Pietro Cicuta}
\address{Cavendish Laboratory, University of Cambridge, U.~K.}

\begin{abstract}
We investigate experimentally the relationship between local structure and dynamical arrest in a quasi-2d colloidal model system which approximates hard discs. We introduce polydispersity to the system to suppress crystallisation. Upon compression, the increase in structural relaxation time is accompanied by the emergence of local hexagonal symmetry. Examining the dynamical heterogeneity of the system, we identify three types of motion : ``zero-dimensional'' corresponding to $\beta$-relaxation, ``one-dimensional'' or stringlike motion and ``two-dimensional'' motion. The dynamic heterogeneity is \emph{correlated} with the local order, that is to say locally hexagonal regions are more likely to be dynamically slow. However we find that lengthscales corresponding to dynamic heterogeneity and local structure do not appear to scale together approaching the glass transition.
\end{abstract}

\maketitle

\section{Introduction}

Identifying the nature of the glass transition is one of the major challenges in condensed matter physics. A number of theoretical approaches have been advanced~\cite{cavagna2009,berthier2011}, but obtaining data which enables discrimination between these is challenging, not least because the timescales for equilibration would diverge at any transition. Broadly there are two schools of thought :  either the glass transition is connected to an underlying thermodynamic singularity to an ``ideal glass'' or its origin is predominantly dynamical. Some theories of the ``thermodynamic'' standpoint envisage a transition to a state rich in certain geometric motifs ~\cite{tarjus2005}, somewhat reminiscent of crystallisation, however the local structures formed do not tile Euclidean space ~\cite{frank1952}.

Here we shall be concerned with supercooled liquids whose relaxation we can identify on the experimental timescale rather than solid (nonequilbrium) glasses. Among the more striking observations in supercooled liquids is that of dynamic heterogeneity where spatio-temporal fluctuations in dynamics indicate that some parts of the system are --- transiently --- more solid-like that others. Originally identified in computer simulation ~\cite{hurley1995,perera1996} and indirectly observed in molecular experiment ~\cite{schmidtrohr1991,cicerone1995}, dynamic heterogeneity was later directly observed in colloidal experiments ~\cite{weeks2001,kegel2001,rice01}. The latter technique, which we employ here, is a powerful means of investigating the local properties of glassforming systems ~\cite{hunter2012,yunker2014,ivlev}.

Regarding the nature of the glass transition, strong evidence has been presented that the super-Arrhenius increase in structural relaxation times manifested in supercooled liquids necessitates some kind of change in structure \cite{montanari2006}. Although two-point measures such as the radial distribution function $g(r)$ and its reciprocal space counterpart the static structure factor show little change on approach to the glass transition ~\cite{berthier2009,royall2014physrep}, in recent years through use of higher-order structural measures, considerable evidence in support of a change in structure approaching the glass transition has emerged ~\cite{royall2014physrep,dzugutov2002,shintani2006,coslovich2007,karmakar2009,mazoyer2011,mosayebi2010,royall2008,sausset2010,sausset2010pre,charbonneau2012,leocmach2012,malins2013jcp,malins2013fara,eckmann2008,royall2014}. Many of these studies have shown that dynamically slow regions (i.e. those which are more ``solid'') are correlated with certain structural motifs. However correlations between structure and dynamics do not necessarily demonstrate that this change in structure is the \emph{cause} of the slow dynamics \cite{charbonneau2012}. Support for the thermodynamic viewpoint which relates structure and dynamic arrest would come from growing structure correlation lengths, especially if these grew concurrently with lengthscales related to dynamic heterogeneity. Different means to obtain a structural length have been investigated in detail~\cite{shintani2006,leocmach2012,malins2013jcp,malins2013fara,pedersen2010,royall2008,kawasaki2007,kawasaki2010jpcm,hocky2012,royall2014}. These and other approaches to determine lengthscales, both static and dynamic, have recently been reviewed by Karmakar \emph{et al.} ~\cite{karmakar2014} to which we direct the interested reader. In summary, there is as yet no conclusive outcome. Although a number of groups, particularly in dimension $d=3$, have found that dynamic lengths (usually manifested in the so-called four-point dynamical correlation length $\xi_4$ ~\cite{lacevic2003}) increase much faster than structural lengths approaching dynamical arrest ~\cite{karmakar2009,malins2013jcp,malins2013fara,charbonneau2012,royall2014,hocky2012,royall2014physrep,dunleavy2012}, some have found that $\xi_4$ \emph{does} scale with structural lengths ~\cite{mosayebi2010,sausset2010,sausset2010pre,kawasaki2007,kawasaki2010jpcm}.

In two dimensions, the situation with local structural motifs in the liquid is rather special. This is because, unlike the five-fold symmetric icosahedra and variants encountered in 3d, in 2d simple liquids the local structural motif is the hexagon which \emph{does} tile the (Euclidean) space. Thus there is no inherent geometric frustration in two dimensions. As a consequence it is hard to prepare supercooled liquids and indeed the freezing transition of hard discs has a very different nature in 2d compared to 3d in that it is only weakly first order \cite{bernard2011,engel2013}. Thus to form a supercooled liquid which is stable on all but the shortest timescales one must introduce frustration, either by curving space \cite{sausset2010,sausset2010pre,irvine2012}, introducing many-body interactions which suppress the inherent hexagonal local symmetry ~\cite{shintani2006} or by using multicomponent or polydisperse systems \cite{dunleavy2012,kawasaki2010jpcm,kawasaki2007,yunker2009,watanabe2008,candelier2010}. Unlike their 3d counterparts, 2d systems have been shown to exhibit long-ranged hexagonal order which may be treated in a simple way with a two-state model \cite{langer2013}. However, questions remain concerning the case of 2d systems. In particular, the range of static correlation lengths implied in a thermodynamic viewpoint for the glass transition is intimately related to the degree of frustration. In particular more strongly frustrated systems have much shorter structural correlation lengths \cite{sausset2010,sausset2010pre,kawasaki2007,dunleavy2012}. This opens questions about how closely large structural correlation lengths \emph{of hexagonal order} might be related to crystallisation. Furthermore, some studies indicate that dynamic correlation lengths may be decoupled from structural lengths especially in the case of higher polydispersity (strong frustration) \cite{dunleavy2012}.

Here we use a quasi-2d model colloidal system to investigate the role of local structure in 2d supercooled liquids. 
We have conducted experiments by optical microscopy, to image structure and dynamics in ``real'' space; we have chosen to work with mixtures of colloidal particles around a few micrometers in diameter, which are large enough not just to resolve each one individually and know its position to high precision, but also to assign its size and thus be able to check against size segregation. A limitation of this choice is that it is challenging to resolve equilibration timescales for relaxation of collective structures, which would be too long-lived. 
Previous experimental work in aging (non-equilibrium) systems has shown an intriguing connection between local hexagonal symmetry and dynamics~\cite{yunker2009}. Our interest here is in the supercooled liquid where we focus on local structure and probe the question of increasing lengthscales approaching the glass transition.

This paper is organised as follows. In section~\ref{sectionMaterialsAndMethods} we describe our methodology. In the results (section~\ref{sectionResults}) we first discuss the dynamical behaviour of the system, before moving to the local structure. We then consider correlations between structure and dynamics and the emergence of lengthscales in the system as it approaches arrest. In the final section ~\ref{sectionConclusions} we offer our conclusions.

\section{Materials and Methods}
\label{sectionMaterialsAndMethods}

\subsection{Colloidal dispersions preparation}
Our quasi-2d model system consists of monolayers of polydisperse hard-sphere colloidal particles in water. Dispersions with 2\% by weight of particles in Milli-Q water are prepared with three kinds of silica particles from Bangs Laboratories (species A with diameter $\sigma_A = 3.01\,\mu$m, species B with $\sigma_B = 3.47\,\mu$m and species C with $\sigma_C = 3.93\,\mu$m). To obtain samples with various polydispersity, several dispersions are prepared by mixing particles with different combinations (species A and B, species A and C, species A and B and C). In our system, the gravitational length scaled by the particle size is $\lambda_g/\sigma=0.00976$ for the smallest particles ($\sigma_A =3.01$ $\mu$m) and correspondingly smaller for the larger particles. We thus conclude that thermal fluctuations out of plane are small relative to the particle size and are henceforth neglected, so we consider our experiments to behave as a 2d system. We neglect any nonaddivity effects due to the different particle sizes as these are expected to be very slight for our parameters ~\cite{assoud2010}.

\subsection{Assembling the particle monolayer}
\begin{figure*}[t]
\begin{center}
\includegraphics[width=16cm]{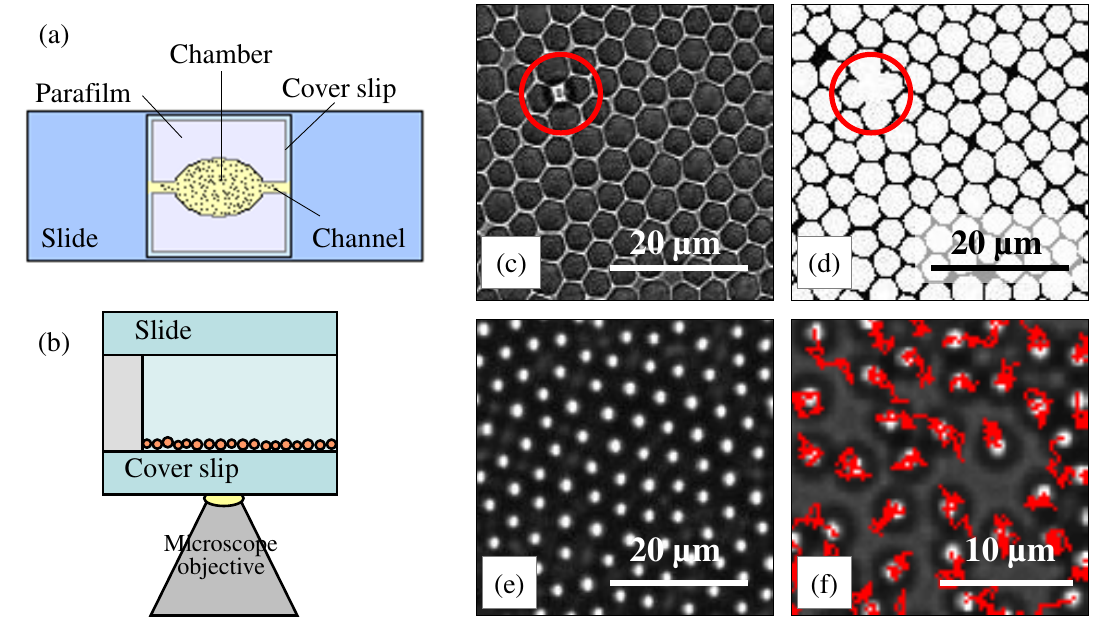}
\caption{Experiments allow tracking and identifying size of particles in dense 2d systems. (a), (b) Schematic view of the experimental set-up. The sample cell showed in (a) is built with a microscope slide and a cover slip stuck together with two pieces of Parafilm  such  that a chamber with two open channels is formed. This is filled with the particle dispersion and sealed. Samples are observed with an inverted microscope, so that particles sediment to form a monolayer (b). (c) Crop of one image acquired focusing the particle equator. The particle contour is clearly visible, as well as the inter-particles spaces. When a second layer is present, particles in the second plane have a clear optical signature, as highlighted by the red circle. (d) The same image segmented and thresholded; white pixels reproduce the particle shape and black pixels are the interstices. When one particle in an extra-layer is present, the black pixels of the interstices are not visible as shown in the red circle. (e) Crop of an image acquired focusing on the particle poles so that particles appear as bright dots in a dark background, greatly facilitating tracking. (f) The position of each particle is tracked during time and the particle trajectories are visualized in red.   }
\label{figSampleImages}
\end{center}
\end{figure*}

Quasi-2d monolayers are prepared in a cell built from a microscope slide and a cover slip. First microscope slides and cover slips are cleaned with water and detergent to make the surface hydrophobic and to prevent particles from sticking to it. The cell is prepared by sticking together a cover slip and a microscope slide with two pieces of Parafilm slightly heated on a heater plate and cut as shown in Fig.~\ref{figSampleImages}(a). In this way a chamber with two open channels is formed between the microscope slide and the cover slip. The chamber is filled injecting the particle dispersion with a micropipette in the proximity of one of the two channels. After this, the channels are sealed with ultra high vacuum grease.
Samples are observed as shown in Fig.~\ref{figSampleImages}(b). In this way, particles settle on the cover slip due to gravity forming a planar layer. By injecting dispersions prepared by mixing particles of species A, B and C with different compositions and concentrations, we obtain monolayers with various polydispersity and various area fractions.

\subsection{Imaging}
A Leica DMI6000 Inverted Microscope, with a 63x~HCX PL FLUOTAR oil-immersion objective, is used in brightfield mode to visualize the monolayers. For each sample, two acquisition methods are used. First, one image with size 512x512\,pixel (1 pixel corresponding to $0.24\,\mu$m) is recorded on a camera (Leica DFC350-FX) focusing on the particles' equators. Figure~\ref{figSampleImages}(c) is a crop of an image acquired with this method. The particle contours are clearly visible, as well as the interstices, so that we can be sure particles are dispersed in a monolayer. When one particle in a second layer is present, it has a clear optical signature, as highlighted by the red circle in Fig.~\ref{figSampleImages}(c). Here we take care only to analyse data where the particles are strictly in a monolayer.
With the second acquisition method, focus is on the particle poles and particles appear as bright dots in a dark background. Acquired images have size 512x512 pixels and a cropped image is shown in Fig.~\ref{figSampleImages}(e). With this configuration, series of 250, 500 or 1000 images are recorded at 1\,fps, so that the sample evolution in time is monitored for an interval that goes from 250\,s to about 17\,minutes, depending on the experiment. With both acquisition methods, in a given image between 950 and 1350 particles are observed, depending on the sample area fraction.

\subsection{Particle tracking}
In Fig.~\ref{figSampleImages}(e), we see that, with the second acquisition method, particles appear as bright dots in a dark background. Using such images it is possible to follow the evolution of the position $\vec{x}_i(t)$ of the center of each particle in time, by tracking the position of the white spots. These series of images are analysed using software developed in house for correlation filtering and sub-pixel resolution of particle positions~\cite{cicuta07c, cicuta09a}. The software recognizes the particle positions in each frame of a series,  identified as a red point in Fig.~\ref{figSampleImages}(f), where complete trajectories are overlapped to the initial image of the series. In certain samples some collective drift is observed. The drift is removed in the data processing so that the position of the i$^{th}$ particle at the time $t$ is

\begin{equation}
\vec{X}_i\left(t \right) = \vec{x}_i\left(t \right)-\langle\vec{x}_i(t) \rangle_i+\langle\vec{x}_i(0) \rangle_i
\label{eq:nodrift}
\end{equation}

\noindent where $\vec{x}_i(0)$ is the i$^{th}$ particle initial position and $\langle ... \rangle_{i}$ is the average on all the particles tracked in  one image. In the rest of this paper, trajectories and all the related quantities are given after drift subtraction.

\subsection{Characterization}
Images taken with the first acquisition method [Fig.~\ref{figSampleImages}(c)] can be converted into black and white images [Fig.~\ref{figSampleImages}(d)] where white pixels reproduce the particle's shape and black pixels are the interstices.
Black and white images are used to quantify the sample packing fraction $\phi$ and polydispersity $s$. We define the area fraction
as the ratio between the number of white pixels and the number of pixels of an image. As shown in Fig.~\ref{figSampleImages}(d), the different particle sizes present in a monolayer are clearly distinguished and this allows us to identify the sample polydispersity  as defined by:
\begin{equation}
s =\frac{\sqrt{\langle \sigma^2 \rangle_\mathrm{species} - \langle \sigma \rangle_\mathrm{species}^2}}{ \langle \sigma \rangle_\mathrm{species}} ,
\label{polydisp}
\end{equation}
where $\langle \sigma \rangle_\mathrm{species}$ is the average diameter of the different species A, B and C present in the sample. Forty samples with particle packing fractions between 0.610 and 0.822 and polydispersity between $s = $7\% and 13\% are considered in our dataset.

The particles are expected to behave as strongly screened charged colloids which are a reasonable model for hard spheres (and here hard discs)~\cite{royall2013myth}. We estimate the interparticle interactions as follows. An approximation to the upper bound of the (effective) colloid charge $Z$ is to set $Z \lambda_B/\sigma=6$ ~\cite{royall2013myth}, where $\sigma$ is the particle diameter. For our parameters this yields an effective charge number of $Z=2.97 \times 10^4$. We suppose the ionic strength is dominated by the counterions and that these are confined to a layer whose height is equal to the mean particle diameter $3.47$\,$\mu$m. In the solvent the ionic strength is then $2.58$\,mMol which corresponds to a Debye length of $\kappa^{-1}=270$\,nm. The dimensionless inverse Debye length $\kappa \sigma=12.8$ (taking $\sigma=3.47$\,$\mu$m). For these parameters reasonably hard-disc like behaviour is expected.

\section{Results}
\label{sectionResults}

\subsection{Dynamics}

\begin{figure}[!t]
\begin{center}
\includegraphics[width=9cm]{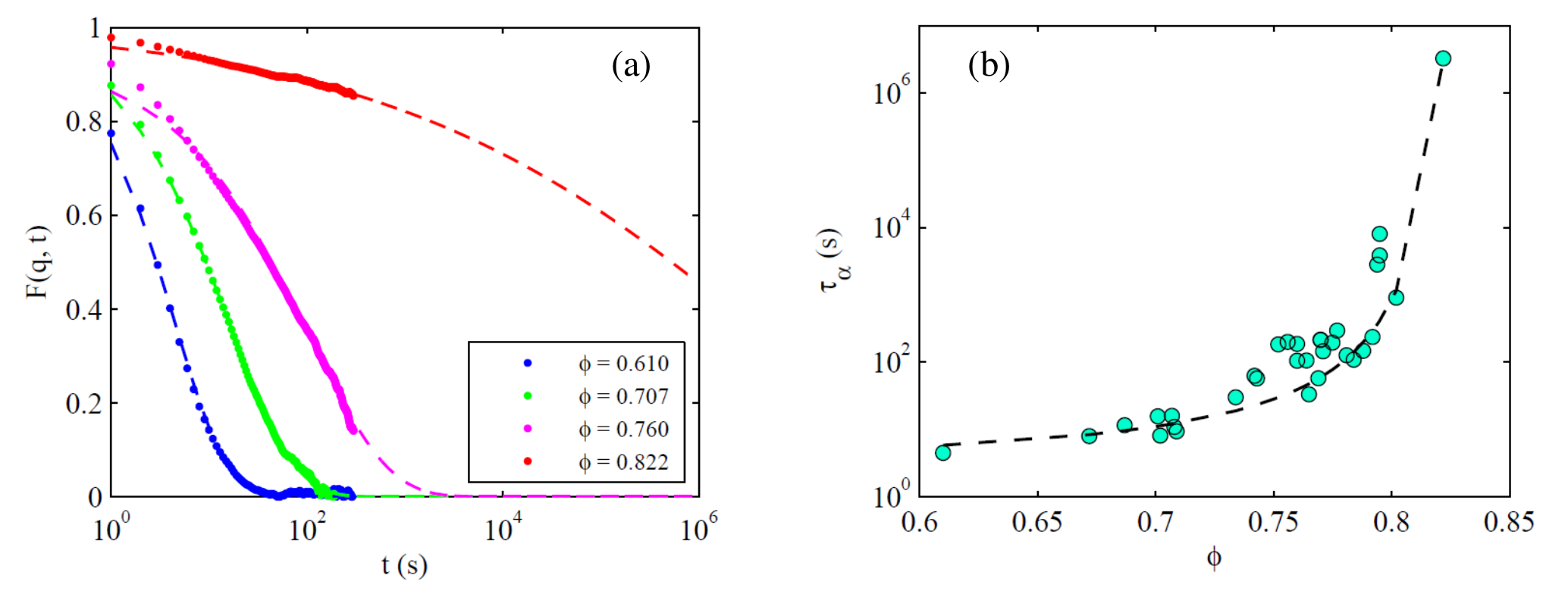}
\caption{The system relaxation time can be obtained for most samples. (a) Intermediate scattering function versus time for four representative samples:  $\phi$ = 0.610 and $s$ = 10\%;  $\phi$ = 0.707 and $s$ = 11\%;  $\phi$ = 0.760 and $s$ = 11\%;  $\phi$ = 0.822 and $s$ = 11\%. Experimental data are fitted with $F(q,t)=C\exp [(-t/\tau_{\alpha})^\beta]$ (dashed lines). (b) System relaxation time $\tau_{\alpha}$, obtained from the fit of $F(q,t)$, versus the area fraction $\phi$. The dashed line is a fit according to the Vogel-Fulcher-Tamman form Eq.~\ref{eqVFT}.}
\label{figScattTau}
\end{center}
\end{figure}

To characterise the dynamical behaviour of the sample we determine the intermediate scattering function (ISF) and the mean-squared displacement. The former quantity we use to characterise the global dynamics of the system, the latter provides a convenient measure of the dynamics at the particle level. We begin by discussing the ISF and the overall dynamics.
To calculate the ISF, we Fourier transform images with eight periodic neighbours in a square lattice. We then consider a ring with internal radius 41-7 pixels and external radius 41+7 pixels, where 41 pixels is the position of the first peak of the structure factor. We then carried out a pixel by pixel autocorrelation in time, and averaged over all these pixels. These we plot in Fig.~\ref{figScattTau}(a). At high area fractions, our ISFs do not fully decay (to get full decay over longer timescales, one needs much smaller particles \cite{brambilla2009}, which would prevent the other measurements we do in this work). We fit throughout with a stretched exponential form
\begin{equation}
F(q,t)=C\exp \left( \frac{-t}{\tau_\alpha} \right)^\beta,
\label{eqKWW}
\end{equation}

\noindent where $C \leq 1$ and $\beta \leq 1$  and $\tau_\alpha$ is the structural relaxation time.
We plot these structural relaxation times as a function of area fraction $\phi$ in the ``Angell plot'' in Fig. \ref{figScattTau}(b). The experimental values are then fitted with a Vogel-Fulcher-Tamman ~\cite{cavagna2009,berthier2011} form
\begin{equation}
\tau_{\alpha}(\phi)=\tau_0 \exp\left( \frac{D}{\phi_0-\phi} \right),
\label{eqVFT}
\end{equation}
\noindent where we obtain $\phi_0 = 0.838$ for the divergence of the structural relaxation time and $D = 0.225$ for the fragility parameter. These values are comparable to those found in the computer simulation literature ~\cite{kawasaki2007,dunleavy2012}. Thus our system exhibits the slow dynamics typical of a model glassformer.

\subsection{Polydispersity and mean-squared displacement}
\begin{figure*}[!t]
\begin{center}
\includegraphics[width=12cm]{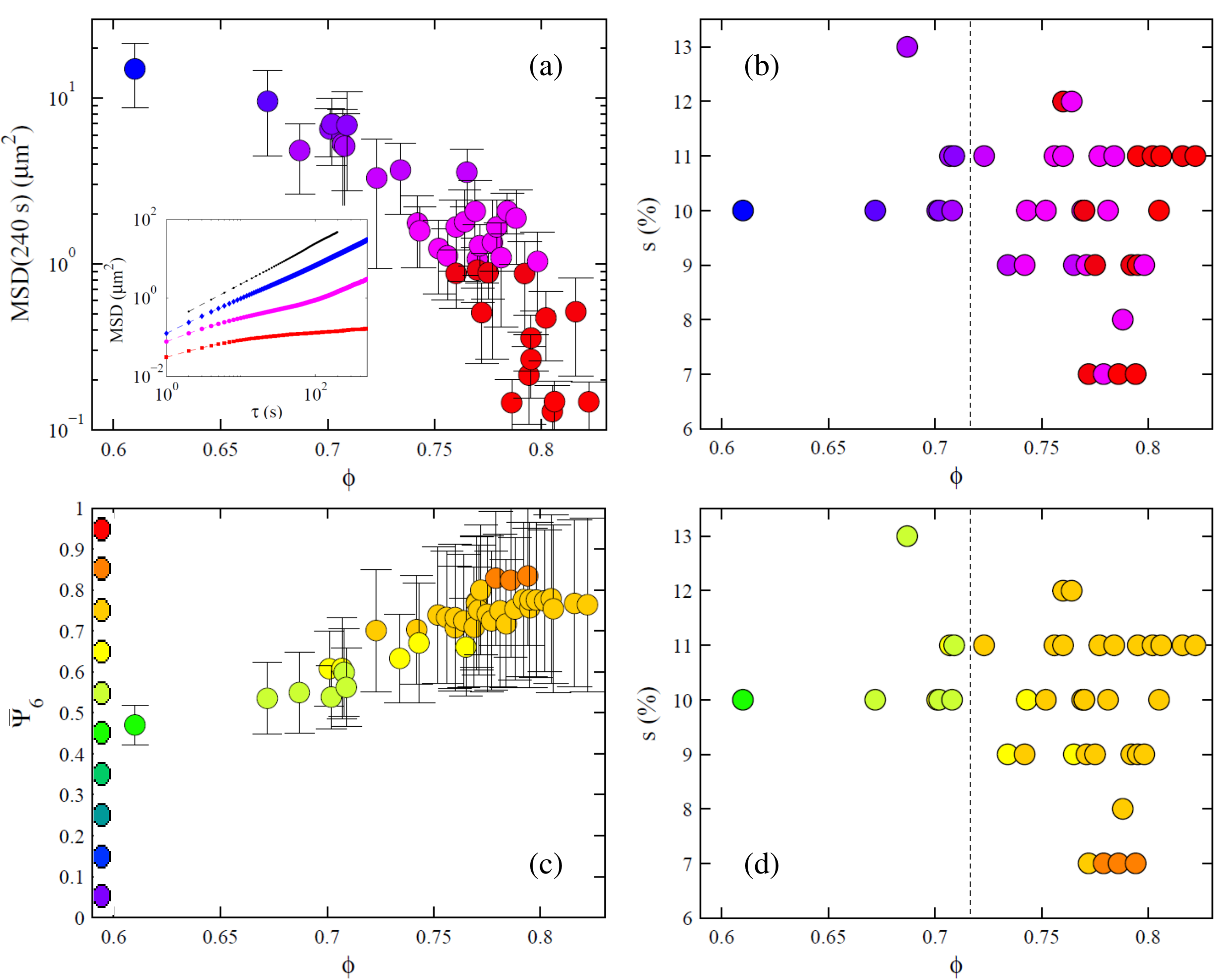}
\caption{With increasing density, dynamics slows down and order sets in; this is quantified here and resolved versus the sample polydispersity. {(a) MSD for $\tau=240$ s as a function of $\phi$. Values are taken from the $\langle\vec{X}^2(\tau)\rangle$ of each sample. Error bars represent the standard deviation of the distributions. Inset: mean-squared displacement vs  $\tau$ for four representative samples:  ($\diamondsuit$) $\phi$ = 0.610 and $s$ = 10\%;  ($\circ\circ$) $\phi$ = 0.760; $s$ = 11\%; $(\Box)$: $\phi$ = 0.822 and $s$ = 11\%); black dots represent the MSD of a dilute state point ($\phi$ = 0.168). (b) State diagram $s$ vs $\phi$, describing the sample dynamics. To define the dynamics of each sample we consider the sample mean-squared displacement MSD for $\tau=240$\,s showed in (a). (c)~Sample averaged $\overline{\Psi}_6$ as a function of $\phi$ for the forty investigated samples. Error bars represent the standard deviation of the ${\Psi}_6$ distributions; note the increase in $\overline{\Psi}_6$ with density, and also more interestingly the emergence of heterogeneity. (d) State diagram $s$ vs $\phi$, showing the sample averaged $\overline{\Psi}_6$. In (b) and (d), the dashed  lines highlight the region $0.684 <\phi<0.704$, where the liquid phase and the solid phase coexist in monodisperse two dimensional colloidal systems~\cite{rice00}. } 
\label{figPD} }
\end{center}
\end{figure*}

The samples analysed for this work are plotted in a polydispersity $s$ vs area fraction $\phi$ state diagram describing the sample dynamics [Fig.~\ref{figPD} (a)].
The dynamics of the $i^{th}$ particle can be characterized by measuring its mean-squared displacement, defined as the time average:

\begin{equation}
\langle \vec{X}_i^2\left(\tau \right)\rangle=\langle (   \vec{X}_i(t+\tau) - \vec{X}_i(t) )^2  \rangle_t.
\label{eq:iMSD}
\end{equation}

\noindent This quantity characterizes the kind of motion exhibited by the particle, allowing one to distinguish between free Brownian motion, where mean-squared displacement (MSD) is linear in $\tau$, from confined motion where the square displacement grows sublinearly or even plateaus at large $\tau$. By computing the sample MSD as an average of $\langle \vec{X}_i^2\left(\tau \right)\rangle$ on all the particles of one image, we can investigate the global dynamics of each sample. Representative mean-squared displacement data as a function of $\tau$ are shown in the inset of Fig.~\ref{figPD} (a) for different sample packing fractions. For the monodisperse sample of particles with $\sigma_B = 3.93\,\mu$m, at low density ($\phi$ = 0.168), the  MSD is linear in $\tau$, as expected in diluted samples; in two dimensions one has ${\langle \vec{X}^2\left(\tau \right)\rangle=4D\tau}$ (where $D$ is the diffusion coefficient), and by fitting that dataset  $D=0.6*10^{-13}\,$m$^2$/s. The Stokes-Einstein coefficient, defined as $D_\mathrm{SE}=k_BT/ 3 \pi \eta \sigma_B$ (with $k_B$ Boltzmann constant, $T$ temperature of the system and $\eta$ water viscosity), for our particles is $D_\mathrm{SE} = 1.16 \times 10^{-13}\,$m$^2$/s. If particles diffuse close to a surface, as in our case, the diffusion is slowed down by approximately a factor of up to 3 compared to bulk diffusion~\cite{brenner83}, so the value of $D$ we find from the fit is in agreement with the expected value. The MSD for a sample with $\phi$ = 0.610 is becoming sublinear. For the sample with $\phi$ = 0.760  the sublinearity is evident and, increasing again the packing fraction ($\phi$ = 0.822), the MSD shows a plateau.

To discriminate between different dynamics and build the $s$ vs $\phi$ state diagram presented in Fig.~\ref{figPD}(b), we use the value of MSD at the time lag $\tau=240$\,s (i.e. $\langle\vec{X}^2(240\,\mathrm{s})\rangle$) which is shown in Fig.~\ref{figPD}(a) as a function of the packing fraction. Error bars represent the standard deviation of the $\langle\vec{X}_i^2(240\,\mathrm{s})\rangle$ distributions (note: this distribution widens as heterogeneity emerges, and then narrows at high concentration).
In Fig.~\ref{figPD}(a) marker color represents the  values of $\langle\vec{X}^2(240\,\mathrm{s})\rangle$, consistently with the colors used in Fig.~\ref{figPD}(b). It is evident that the dynamics slows down for samples with higher packing fraction.

Moreover, looking at Fig.~\ref{figPD}(b), samples with a smaller $s$ (7\%) seem to reach the arrested state at lower $\phi$ and this is in agreement with the fact that polydispersity delays the slowing down which might be related to the fact that random close packing occurs at a higher $\phi$ in the case of higher polydispersity~\cite{torquato2010}. It has been estimated~\cite{rice00} that for a monodisperse two dimensional colloidal system the liquid-to-solid transition happens in the region 0.684 $<\phi<$ 0.704, that is the range of $\phi$ where the liquid phase and the solid phase coexist. This region is highlighted in Fig.~\ref{figPD}(b) and Fig.~\ref{figPD}(d) by the two dashed black lines and it is clear that in our polydisperse system, the dynamics is  not totally slowed down even beyond these values of $\phi$.

A second $s$ vs $\phi$ state diagram is shown in Fig.~\ref{figPD}(d). Here data are coloured according to the mean structural characteristic in the sample [colormap in Fig.~\ref{figPD}(c)]. In disordered granular systems~\cite{watanabe2008} and colloidal glasses~\cite{leocmach2012}
long-lived (relative to $\tau_\alpha$) medium-range crystalline regions have been found. To investigate the local structure
we use the bond-orientational order parameter defined for each particle $j$ as

\begin{equation}
\ \psi_6^j=\Big| \frac{1}{n_j}\sum_k\exp{(i6\theta_{jk})}\Big|,
\label{eq:psi6}
\end{equation}

\noindent where the sum runs over the $n_j$ nearest neighbours of the particle and $\theta_{jk}$ is the angle between $\vec{r}_{k}-\vec{r}_{j}$ and a fixed arbitrary axis. Nearest neighbours are identified by a Voronoi construction. Hereafter, we use $\Psi_6^j=\langle \psi_6^j \rangle_t$, where the time average of the order parameter is computed on a given number of frames from the beginning of the acquisition. When the particle $j$ is
in a locally hexagonal configuration  $\Psi_6^j=$ 1. The more $\Psi_6^j$ tends to zero, the more the particle is in a disordered region. If $\Psi_6^j$ is averaged over all the particles of each sample, we have the parameter $\overline{\Psi}_6=\langle \Psi_6^j \rangle_j$ to define the degree of order in a sample. In Fig.~\ref{figPD}(c) $\overline{\Psi}_6$ is plotted as a function of $\phi$; error bars represent the standard deviation of the ${\Psi}_6$ distribution and the same color code is used for the state diagram in Fig.~\ref{figPD}(d). The formation of order is favoured upon compression, and the value of $\overline{\Psi}_6$ for the range of sample packing fraction used in our experiments, increases from around 0.5 to 0.85, indicating that none of the samples is crystalline.

By comparing the two state diagrams [Figs.~\ref{figPD}(b) and (d)] the slowing down and the formation of order are more evident for high sample packing fraction and their appearance is slowed by polydispersity. According to this observation slowing down and order are related and this is consistent with previous observations in computer simulation~\cite{kawasaki2007} and experiment~\cite{watanabe2008,williams2014jcp}.

\subsection{Particle trajectories and dynamic heterogeneity}

\begin{figure*}[t]
\begin{center}
\includegraphics[width=16cm]{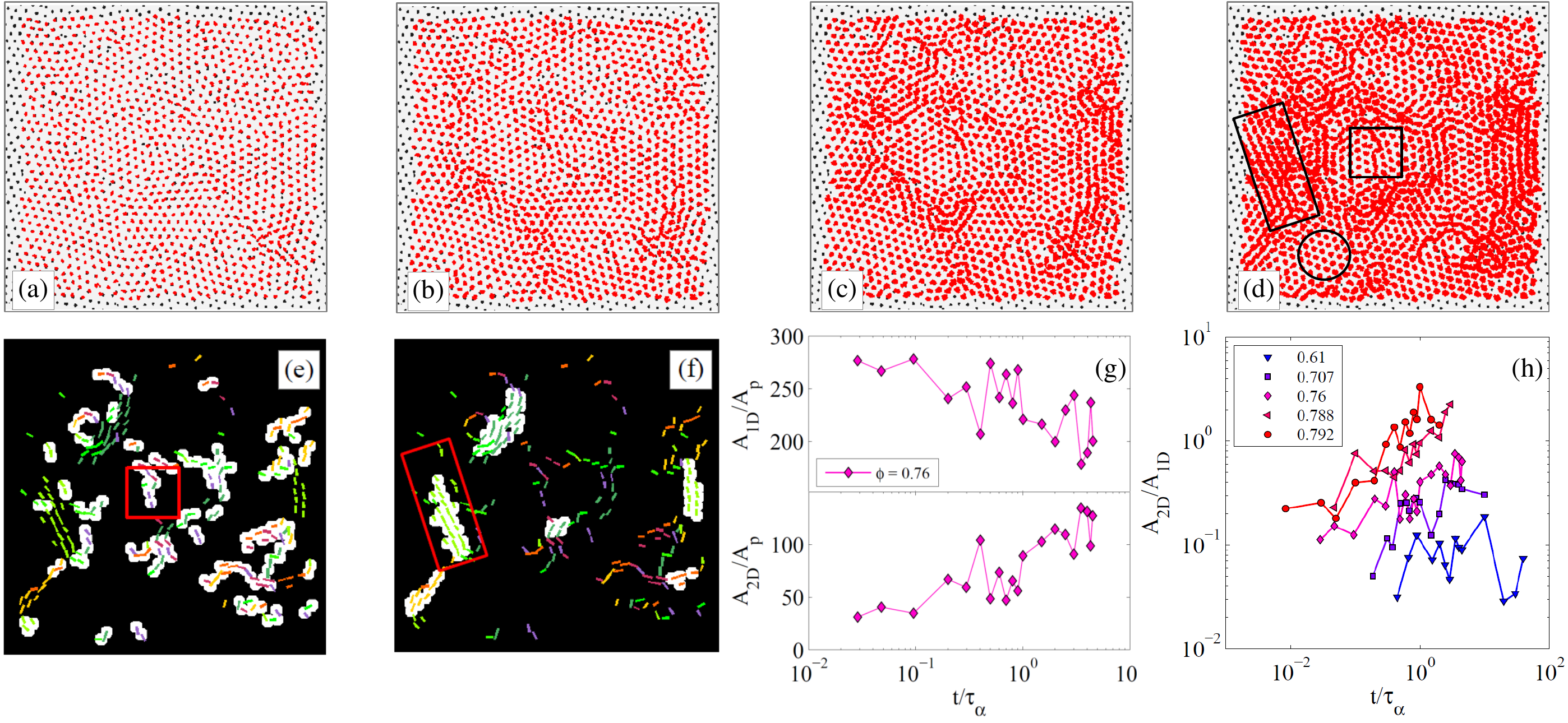}
\caption{Particle trajectories are uniform at short time intervals, but spatially heterogeneous over longer intervals. Shown here are tracks over 0.3$\tau_\alpha$~(a), $\tau_\alpha$~(b), 2.3$\tau_\alpha$~(c), 4.3$\tau_\alpha$~(d), measured from the beginning of the acquisition, for the sample with $\phi= 0.760$ and $s = 11\%$. Trajectories are overlapped to the initial image of the series (black spots represent the particle centers). Particles on the border are not considered in our analysis. The shapes in panel (d) highlight regions of interesting collective motion, which are discussed in the text. At sufficiently long time intervals, the particles that have high absolute displacement, and share a common direction, cluster close to each other; this is shown in panel (e), highlighting the particle clusters of size 2 and 3, and panel (f) highlighting the clusters of 4 and particles and above.  (e,f) are derived from the same waiting time and sample shown in (d). As discussed in the text, we see that high mobility string-like regions emerge in (e), and two-dimensional raft-like regions emerge in (f) (an example of each is highlighted by a rectangle). The area occupied by these thresholded fractions (the clusters) depends on the waiting time: the 1d string-fraction falls, while the 2d raft-like increases with  time interval (g).  The ratio of these fractions has clear trends with the particle density (h).\label{figTrajectTime}
}
\end{center}
\end{figure*}

The appearance of confined motion is not homogeneous in the sample, but rather  heterogeneity in the dynamics is visible, as already observed in several disordered granular~\cite{watanabe2008,candelier2010}
and colloidal~\cite{rice01,weeks2001,kegel2001,leocmach2012} systems. With our experiments, we can investigate the time-evolution of the heterogeneity. We consider the sample with $\phi = 0.760$ and $s = 11\%$, and in Fig.~\ref{figTrajectTime} we present the particle trajectories plotted over intervals of 0.3$\tau_\alpha$~(a), $\tau_\alpha$~(b), 2.3$\tau_\alpha$~(c), 4.3$\tau_\alpha$~(d). At short intervals, all the particles exhibit  confined motion and no significant dynamic heterogeneity is visible; at timescales approaching $\tau_\alpha$, dynamic heterogeneity starts to appear and becomes more evident at even longer time intervals. If we consider the path travelled by particles over  4.3$\tau_\alpha$, we can clearly distinguish three different motions: a ``zero-dimensional'' motion (the classical $\beta$-relaxation) of particles with very confined trajectories (as for example in the region highlighted by the  circle), a ``one-dimensional'' (or string-like~\cite{kob1997,kawasaki2013,rice01}) cooperative motion given by ``chains'' of particles moving together (square) following a linear trajectory (not necessarily in the same direction; we see curved or apparently random walks, etc), a ``two-dimensional'' motion given by regions of particles moving together in a preferential direction (rectangle). Although dynamic heterogeneity has been repeatedly observed in experiments and simulations of two dimensional disordered systems, ``two-dimensional'' motion of regions of particles moving together has received relatively little attention. It is clear that the dynamic heterogeneity appears with increasing interval; in particular, the ``zero-dimensional'' motion is the first one to be visible, while the  ``two-dimensional'' motion is the last.

We investigated further the spatial structure of the high-mobility regions, as defined for different waiting times.  For a given sample, the particle displacements at different intervals (between  $0.01\tau_\alpha$ and  $4.5\tau_\alpha$) are considered;  the 20$\%$ of particles with largest displacements are selected. These particles are divided in eight sets, according to the direction of displacement [displacements are represented by color, in Fig.~\ref{figTrajectTime}(e,f)]. When four or more  particles have the same direction of motion and are close to each other, we consider this a ``2d region''; this is a raft, formed of particles moving in the same direction. These regions are highlighted  in white, in Fig.~\ref{figTrajectTime}(f). The remaining of the 20\% fast particles are either isolated, or belong to a dimer or trimer cluster of particles with same direction. We highlight the dimers and trimers in  Fig.~\ref{figTrajectTime}(e). It is clear, looking at the cluster structure in  Fig.~\ref{figTrajectTime}(e), that these dimers and trimers often join together, to form more extended string-like trajectories.  
This  analysis is, empirically, a simple way to isolate strings from rafts:  we define $A_{1D}$ the  total white area in Fig.~\ref{figTrajectTime}(e), and $A_{2D}$ the total white area in Fig.~\ref{figTrajectTime}(f) (both are normalized with the particle area $A_{p}$). 

These areas have a clear dependence on the time interval considered: as a function of $t/\tau_\alpha$,  Fig.~\ref{figTrajectTime}(g) [$\phi$ = 0.760 and $s$ = 11\%, i.e. the same sample of (a), (b), (c) and (d)] shows a decrease of 1d motion and an increase of 2d motion. The trend is robust, and strongest for high density, as shown in Fig.~\ref{figTrajectTime}(h), where $A_{1D}/A_{2D}$ is plotted for five samples with  $\phi = 0.61,0.707, 0.76, 0.788, 0.792)$  (these are the same samples analysed in Figs.~\ref{figPsiMAP} and~\ref{figCorrelations}).  

Regarding the time-evolution of the dimensionality of the motion, it is well-known ~\cite{berthier2009,royall2014physrep} that quantities such as the so-called dynamic susceptibility display a characteristic peak around $\tau_\alpha$ before dying away. This is often interpreted in that dynamically heterogeneous regions have fewer particles at short times, more around $\tau\alpha$ and that the magnitude of the dynamical fluctuations dies away at long times. Our interpretation is that this may be related to the dimensionality in the 0d motion involves little movement at short times, 1d motion is related to smaller number of particles and the 2d motion corresponds to the larger groups of particles involved in dynamic heterogeneity at longer times. Further observations have recently been made indicating that a key source of dynamic heterogeneity at longer times is ``hydrodynamic'' density fluctuations which spread slowly through the supercooled liquid ~\cite{jack2014} which would correspond to our 2d regions.

\subsection{Local structure}
\begin{figure*}[!t]
\begin{center}
\includegraphics[width=16cm]{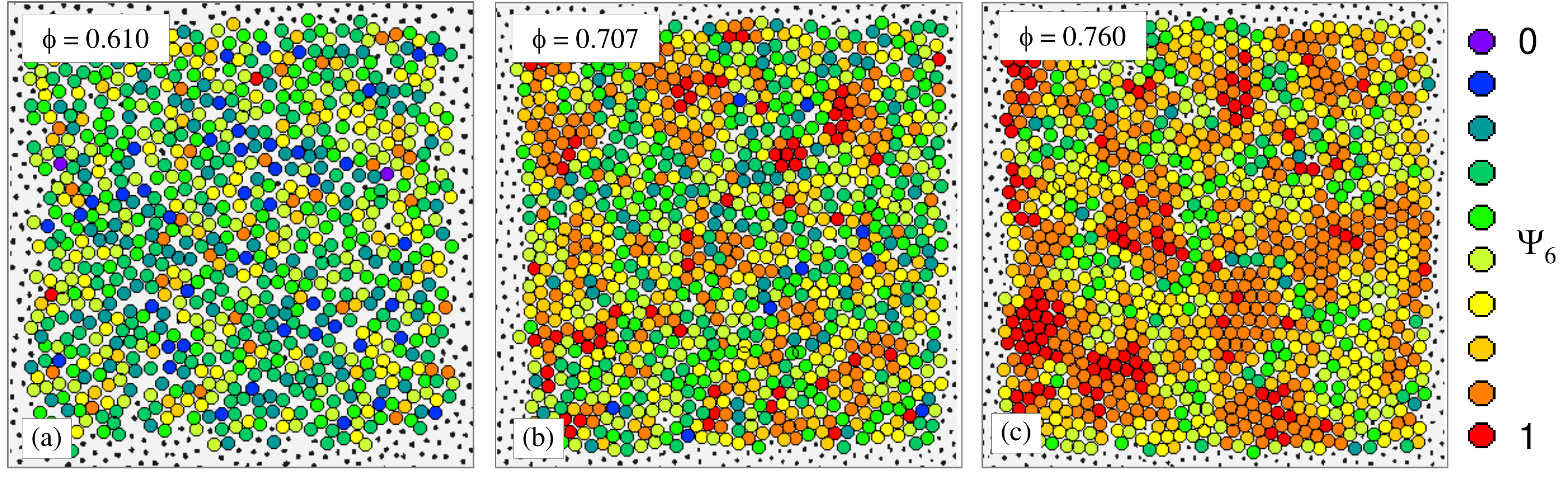}
\caption{The degree of orientational order is uniform at lower density, but develops long range correlations increasing the concentration. Images show  color-maps of the bond-orientational order parameter $\Psi_6$, each circle represents a particle, centered on its initial position, for three samples: (a)  $\phi$ = 0.610 and $s$ = 10\%,  (b)  $\phi$ = 0.707 and $s$ = 11\%,  (c)  $\phi$ = 0.760 and $s$ = 11\%. The $\Psi_6$ is time-averaged over a time interval $\tau_\alpha$, from the beginning of the acquisition. $\Psi_6$ values are divided in ten intervals, each of which is identified by a color as shown in the legend. }
\label{figPsiMAP}
\end{center}
\end{figure*}

To investigate regions with hexagonal order in our samples, we use the bond-orientational order parameter $\Psi_6$. In particular, we calculate the time average of the order parameter $\Psi_6^j=\langle \psi_6^j \rangle_t$
over a time corresponding to $\tau_\alpha$, in order to detect long-lived ordered regions. $\Psi_6$ color-maps can be used to  visualise the presence of ordered and disordered regions in our samples: In Fig.~\ref{figPsiMAP} we present such color-maps of the bond-orientational order parameter $\Psi_6$ per particle for four
samples with comparable polydispersity: (a) sample with $\phi$ = 0.610 and $s$ = 10\%,  (b) sample with $\phi$ = 0.707 and $s$ = 11\%,  (c) sample with $\phi$ = 0.760 and $s$ = 11\%. $\Psi_6$ is time-averaged over an interval $\tau_\alpha$, where $\tau_\alpha$ is 4.5\,s for sample (a), 16\,s for sample (b) and 104\,s for sample (c).
The $\Psi_6$ value of a given particle is represented by a circle of the colour in the key placed on the map with the initial coordinates of the particle. Circles are overlapped to the initial state of the system (black spots are the particles).
In the less concentrated sample [Fig.~\ref{figPsiMAP}(a)] the bond-orientational order parameter is relatively homogeneous and most of the particles have a $\Psi_6$ value between 0.3 and 0.7, with the $\Psi_6$ value distribution centred in 0.5. We can say that the sample is totally disordered and no ordered regions are present. For the sample with $\phi$ = 0.707 [Fig.~\ref{figPsiMAP}(b)] the $\Psi_6$ value distribution is centred in 0.6 and significantly ordered regions with $\Psi_6$ between 0.7 and 0.9 appear (orange disks). The sample structure starts to be quite heterogeneous, since regions of particles with $\Psi_6$ value higher than 0.7 appears. Increasing the area fraction further to $0.760$ [Fig.~\ref{figPsiMAP}(c)], regions with $\Psi_6$ value between 0.7 and 0.9 start to have a considerable size, comprising tens of particles. Regions with $\Psi_6$ between 0.9 and 1 appear. The sample structural heterogeneity is significant.

Note that working with these relatively large colloidal particle sizes allows us to check the size of each particle, and hence the local polydispersity. This is an interesting check, especially on the regions of high angular order, to verify that these are not linked to some chance or induced monodisperse patch. Considering the 21 particles with high $\Psi_6$ (red disks) on the bottom left of Fig.~\ref{figPsiMAP}(c), polydispersity is $s=13\%$; including the adjacent 21 orange disks on their right, $s=12\%$. These are essentially the same as the average over this sample, which is $s=11\%$.

\subsection{Correlation of dynamics and structure}
\label{sectionCorrelationOfDynamicsAndStructure}

\begin{figure*}[t]
\begin{center}
\includegraphics[width=16cm]{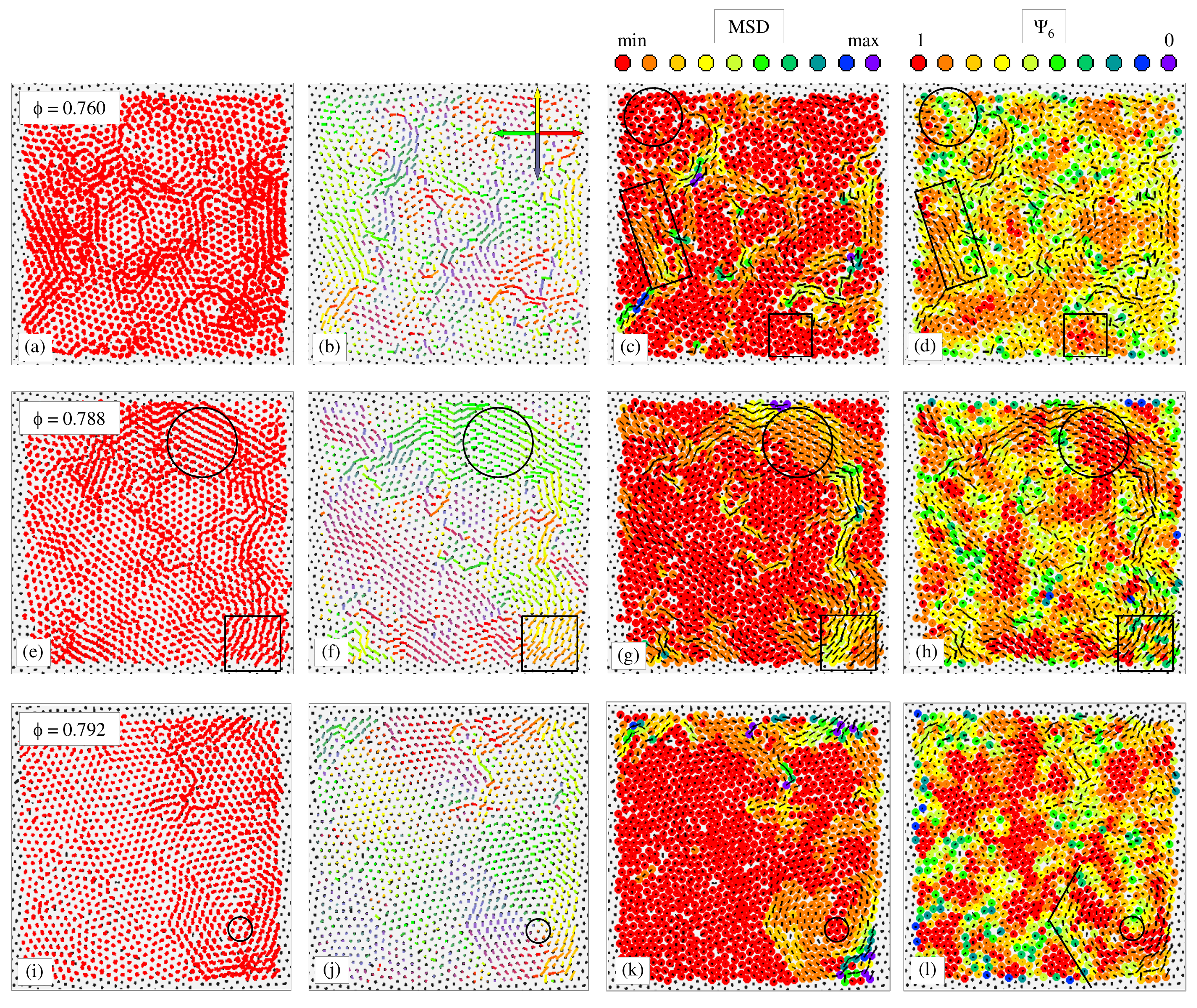}
\caption{There are correlations between the directionality, the absolute mobility and the local order around particles.  Top row: sample with $\phi$ = 0.760 and $s$ = 11\%. Middle row: $\phi$ = 0.788 and $s$ = 8\%. Bottom row:  $\phi$ = 0.792 and $s$ = 9\%.
First column: particle trajectories plotted for a time of 450\,s, which corresponds to $4.3\tau_\alpha$  (a), $t=3\tau_\alpha$  (e), and $t=1.9\tau_\alpha$ (i);
Second column: particle directions of motion represented as displacement vectors joining the position of each particle across an interval of 450\,s and coloured according to the direction (the main four directions are indicated by the coloured arrows on the top-right of the picture);
Third column: colour map of the MSD per particle at the time lag $\tau=450$ s (colour code on the top; intervals are logarithmically spaced);
Right column: colour map of $\Psi_6^j$ time-averaged over 450\,s  (colour code on the top). The displacement vectors showed are overlapped to the colour maps of MSD and $\Psi_6^j$.
}
\label{figCorrelations}
\end{center}
\end{figure*}

By comparing the state diagrams in Fig.~\ref{figPD}(b) and (d), it is clear that the slowing down and the formation of sixfold order are more evident for high area fraction. Investigating the particle trajectories (Fig.~\ref{figTrajectTime}), we have seen the development of dynamic heterogeneity. In the same way, at high $\phi$ long-lived regions with hexagonal structural order are visible in the globally amorphous system. According to this, we may expect that slowing down, dynamic heterogeneity and order are related.

To explore the correlation between these three properties, we first compare the trajectories of each particle in a sample with their MSD and their $\Psi_6$. We show in Fig.~\ref{figCorrelations} three samples, with increasing densities going down the rows.   In the first column,  the particle trajectories are plotted over an interval of 450\,s ($4.3\tau_\alpha$); In the second column, the particle directions of motion are represented as vectors joining the position of each particle across the interval of 450\,s ($4.3\tau_\alpha$),  coloured according to the direction (the Cartesian directions are indicated by the coloured arrows on the top-right of the column). The third column illustrates via  a  colour map the spatial distribution of $\langle\vec{X}^2(450\,\mathrm{s})\rangle$ (colour code at the top of the column, note  intervals are logarithmically spaced). The rightmost column shows the  spatial distribution of  $\Psi_6^j$ time-averaged over the same interval (colour code on the top: red points represent particles with $0.9 \leq\Psi_6^j\leq 1$, violet points represent particles with $0 \leq\Psi_6^j<0.1 $).  The second and third rows contain the matching information, for higher densities (0.788 and 0.792), and here the 450\,s of time interval and averaging correspond to 3$\tau_\alpha$ and $1.9\tau_\alpha$ respectively. In all cases, the time interval considered is $> \tau_\alpha$. In the MSD  and $\Psi_6$ maps in Fig.~\ref{figCorrelations}, the directions of motion are shown overlapped.

Observing the trajectories of the lower concentration sample [panels (c, d)] we see that the three different kinds of motion mentioned above are present: the ``zero-dimensional" motion of particles with very confined trajectories; the ``one-dimensional'' motion given by ``chains'' of particles moving together in a preferential direction; the ``two-dimensional" motion given by regions of particles moving together in a preferential direction. Particles with the ``one-dimensional'' motion have a large MSD and they seem to belong to both ordered and disordered regions. In the region highlighted by the  square, particles have a confined motion, a low MSD and they belong to an ordered region, since their $\Psi_6^j$ is larger than 0.8. But if we consider the particles in the  circle, they have always a confined motion and a low MSD, but they are in a disordered region with $\Psi_6^j$ that goes from 0.3 to 1, depending on the particle. In the  rectangle, particles show a ``two-dimensional'' motion: they are moving together in the up-left direction, but with different MSD and with $\Psi_6^j$ between 0.4 and 1.

Another example of clear  ``two-dimensional'' motion of particles is in the sample with $\phi$ = 0.788 and $s=8$\%, shown in the middle row of Fig.~\ref{figCorrelations}. Particles in the  circle are moving together in the up-left direction, with different MSD and they are part of an ordered region, since their $\Psi_6^j$ is larger than 0.8. Particles in the  square are moving together in the up-right direction, with different MSD but in this case they are part of a disordered region, since their $\Psi_6^j$ is between 0.3 and 1, depending on the particle. From Fig.~\ref{figCorrelations} (f), it seems that the sample is formed by regions of particles moving in a given direction and that the regions together are following an hexagonal path. This observation is evident in the bottom panels of Fig.~\ref{figCorrelations}, showing the sample with $\phi$ = 0.792 and $s$ = 9\%. In the bottom-right part of Fig.~\ref{figCorrelations}(j), particles are moving anticlockwise following an hexagonal path. The center of rotation is indicated by the black circle. In Fig.~\ref{figCorrelations}(k) a net difference in the MSD between particles belonging to the rotating region and particles with a confined motion is visible. One may suppose that the hexagonal path is linked to the hexagonal crystal lattice present in the ordered region, but by looking at the $\Psi_6^j$ colour map in Fig.~\ref{figCorrelations}(l), it is clear that the rotating region is not a unique hexagonal crystal.

\begin{figure*}[!t]
\begin{center}
\includegraphics[width=16cm]{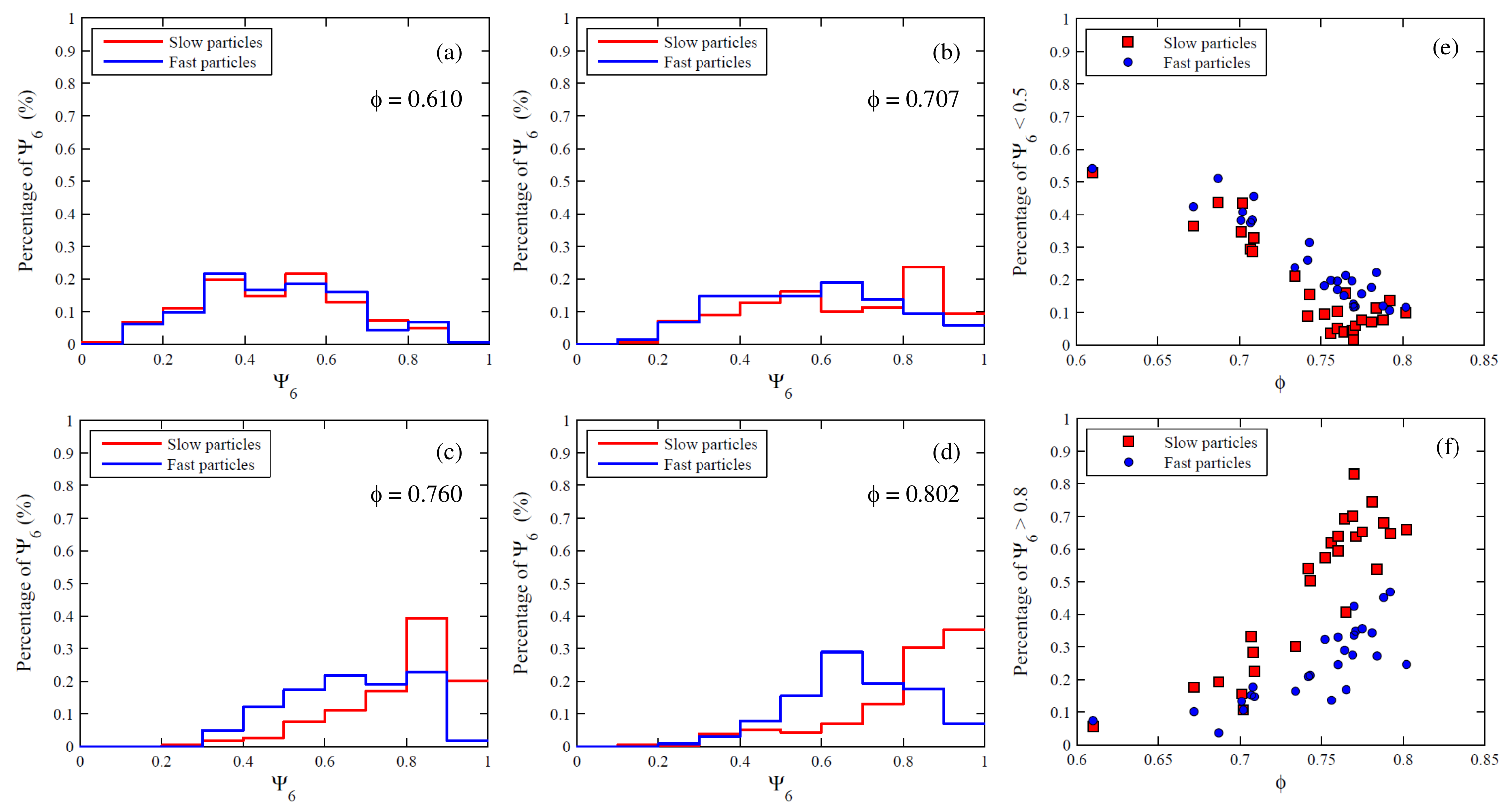}
\caption{Increasing density, structural and dynamic heterogeneity both develop, and they are co-localised. Histograms representing the fraction of particles with a given $\Psi_6$ between the 20\% of the faster particles (blue line) and the fraction of particles with a given $\Psi_6$ between the 20\% of the slowest particles (red line). To discriminate between fast and slow particles, we consider their $\langle\vec{X}_i^2(\tau_\alpha)\rangle$. The four histograms correspond to: (a)  $\phi$ = 0.610 and $s$ = 10\%;  (b)  $\phi$ = 0.707 and $s$ = 11\%; (c)  $\phi$ = 0.760 and $s$ = 11\%; (d)  $\phi$ = 0.802 and $s$ = 11\%. (e) Markers represent ($\bullet$) the fraction of particles with $\Psi_6<0.5$ between the 20\% of the faster particles; ($\blacksquare$) the fraction of particles with $\Psi_6<0.5$ between the 20\% of the slowest particles. Points are plotted versus $\phi$ for the samples for which we can compute $\langle\vec{X}^2(\tau_\alpha)\rangle$. (f) Markers represent ($\bullet$) the fraction of particles with $\Psi_6>0.8$ between the 20\% of the faster particles; ($\blacksquare$) the fraction of particles with $\Psi_6>0.8$ between the 20\% of the slowest particles. Data is plotted versus $\phi$,  for all the samples for which we can compute $\langle\vec{X}^2(\tau_\alpha)\rangle$.
\label{figPSIistogram}}
\end{center}
\end{figure*}

By comparing the $\Psi_6^j$ colour maps with the particle trajectories or the MSD colour maps, some colocalisation between the dynamic heterogeneity and order is visible. To investigate further, we consider the histograms in Fig.~\ref{figPSIistogram} representing the fraction of particles with a given $\Psi_6$ between the 20\% of the faster particles (blue line) and the fraction of particles with a given $\Psi_6$ between the 20\% of the slowest particles (red line). The four histograms correspond to : (a) $\phi$ = 0.610 and $s$ = 10\%;  (b) $\phi$ = 0.707 and $s$ = 11\%;  (c) $\phi$ = 0.760 and $s$ = 11\% and (d) $\phi$ = 0.802 and $s$ = 11\%. At low packing fraction, the fraction of particles with a given $\Psi_6$ is the same for both fast and slow particles and the distributions are centred on $\Psi_6=0.5$. Increasing $\phi$, the fraction of slow particles with $\Psi_6>0.8$ is slightly bigger than that for fast particles. For $\phi$ = 0.760 the fraction of slow particles with $\Psi_6>0.8$ is significantly bigger then for fast particles and in the most deeply supercooled sample almost all the slow particles has a big value of $\Psi_6$, since more then the 70\% have $\Psi_6>0.8$, while the distribution of fast particles is centred in $0.6<\Psi_6<0.7$.  From these histograms it is evident that slow particles belong preferentially to ordered regions and that the slowing down of the dynamics is connected to the formation of long-lived medium-range crystalline order.

Similar results are valid for the samples investigated in our experiments for which we can calculate the MSD at $\tau_\alpha$ ($\langle\vec{X}^2_i(\tau_\alpha)\rangle$). In Fig.~\ref{figPSIistogram}(e) markers represent the fraction of particles with low order, $\Psi_6<0.5$, between the 20\% of the fastest and slowest particles. To discriminate between fast and slow particles, we consider their $\langle\vec{X}^2_i(\tau_\alpha)\rangle$. In Fig.~\ref{figPSIistogram}(f) markers represent  the fraction of particles with high order, $\Psi_6>0.8$, within the 20\% of the fastest and slowest particles for all the $\phi$ considered. The data of Fig.~\ref{figPSIistogram}(e) show that the fraction of fast particles with $\Psi_6<0.5$ is larger than the fraction of slow particles for all the samples, whereas the data of Fig.~\ref{figPSIistogram}(f) show that the fraction of fast particles with $\Psi_6>0.8$ is significantly smaller than the fraction of slow particles for all the considered samples.

\begin{figure}[!t]
\begin{center}
\includegraphics[width=9cm]{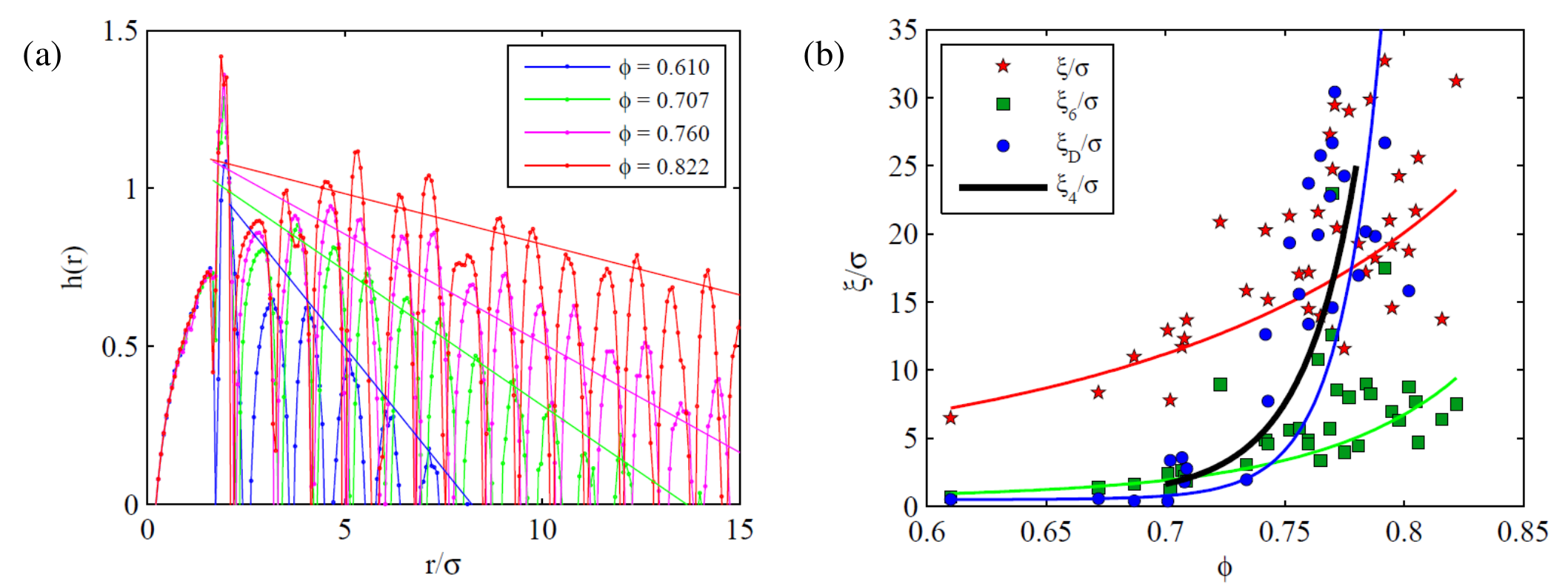}
\caption{Structural and dynamical lengthscales can be measured; when compared to each other, they exhibit different qualitative behaviour as a function of density. (a)$h(r)=\log(r \mid g(r)-1\mid )$ plotted versus $r$ normalized by the average particles diameter $\sigma$ for four representative samples: sample with $\phi$ = 0.610 and $s$ = 10\%; sample with $\phi$ = 0.707 and $s$ = 11\%; sample with $\phi$ = 0.760 and $s$ = 11\%; sample with $\phi$ = 0.822 and $s$ = 11\%. Dashed lines are the linear fit of the $h(r)$ peaks. The inverse of the line slope gives the correlation length $\xi$; these are  plotted in (b) versus $\phi$,  normalized by the average particle diameter $\sigma$. $\xi$ is compared with the $g_6(r)/g(r)$ correlation length $\xi_6$, with the dynamic correlation length $\xi_D$ from $g_D(r)/g(r)$ and with the correlation length $\xi_4$ from literature. The red, blue and green lines in (b) are guides to the eye given by data interpolation.}
\label{figGPsixi}
\end{center}
\end{figure}

\subsection{Structural Lengthscales}

We now turn to a topic which we raised in the introduction: the structural and dynamic lengths approaching the glass transion.
We compute correlation lengths corresponding to two structural quantities. The first is the density-density correlation length. Although this has been relatively little-explored in the context of glassforming systems, it is well-known in the study of critical phenomena ~\cite{onuki}.   We determine the density-density correlation length $\xi$ by fitting $y(r) = Ar^{-1/4}\exp(-r/\xi)$ to  $h(r)=\log(r \mid g(r)-1\mid )$, examples   are shown in Fig.~\ref{figGPsixi}(a). We also obtain the correlation length of local six-fold symmetry in a manner similar to that of Kawasaki and Tanaka~\cite{kawasaki2007}.
We obtain the bond-orientational correlation function $g_6(r) =< \mathrm{Re}[\Psi^*_6(\vec{r'}) \Psi_6(\vec{r'} + \vec{r}) ]>$ by multiplying $\Psi_6$ (complex) of a particle by the $\Psi^*_6$ (complex conjugates) of all the other particles in the sample at a given distance. We then plot $g_6(r)/g(r)$ and fit the peaks with $y(r)= Ax^{-1/4}e^{-r/\xi_6}$.

Our data runs for insufficient time to accurately determine $\xi_4$ (which we take from the literature for a comparable system ~\cite{dunleavy2012}). However we  employ an approach related to that above to obtain a dynamic length. We define
$g_D(r) =\langle  \log [\vec{X}^2(\tau_\alpha, \vec{r'}))]\log  [\vec{X}^2(\tau_\alpha, \vec{r'}+\vec{r})] \rangle$. We then plot $g_D(r)/g(r)$ and fit peaks as for the structural lengths $\xi$ and $\xi_6$.

In Fig.~\ref{figGPsixi}(b) we show these  correlation lengths. 
We find that the density-density correlation length  $\xi$ exceeds the static $\xi_6$, but both exhibit similar weak $\phi$ behaviour.
Interestingly, the dynamic correlation length determined here $\xi_D$ actually exhibits very similar behaviour to $\xi_4$. A direct comparison between our $\xi_D$ and $\xi_4$ determined from the same data would be very helpful, but for now we cautiously note that $\xi_D$ appears to provide a reasonable description of the range of correlated dynamics.  What is evident is that the ``dynamic'' lengthscales grows qualitatively differently from the ``static'' lengths, and we elaborate on the significance of  this  below.

\section{Conclusions}
\label{sectionConclusions}

We have studied the dynamics and its relation to the local structure in a quasi 2d colloidal model system where crystallisation is frustrated by polydispersity. Upon compression local hexagonal order is apparent, and the system becomes more ordered at deeper supercooling. Dynamic heterogeneity is observed and there is some correlation between particles with local hexagonal order and dynamically slow particles. We further investigate the development of the lengthscales associated with both dynamic and structural quantities. Both increase upon supercooling, however dynamic lengthscales seem to increase rather more than do structural lengthscales. Concerning comparable hard disc systems, our findings are consistent with some previous computer simulation work \cite{dunleavy2012} but not with others \cite{sausset2010,sausset2010pre,kawasaki2007,kawasaki2010jpcm}. However we note that our decision to use large colloids means our waiting times are somewhat limited. Thus we cannot rule out that longer equilibration times might change our results and we suggest that this point should be checked carefully in the future. In 3d, a number of studies have found that dynamic lengthscales increase faster than structural lengthscales ~\cite{karmakar2009,malins2013jcp,malins2013fara,charbonneau2012,royall2014,hocky2012} but some suggest that both scale together ~\cite{mosayebi2010,kawasaki2010jpcm}.  We should like to emphasise that the structural correlation length $\xi$ obtained purely from two-point correlations does show an increase comparable to that extracted from the higher-order bond-orientational order parameter. This suggests that it might be possible to investigate structural correlation lengths in certain molecular and atomic systems (for example metallic glassformers and oxides) in which high-precision two-point structural data is available ~\cite{salmon2013}.

The discrepancy between the dynamic lengths we have found and the structural lengths has three possible explanations. Firstly, local structure may be largely unrelated to the slow dynamics as assumed in the dynamic facilitation approach ~\cite{chandler2010}. The second possibility is that the dynamic correlation lengths measured are somehow not representative of the slow dynamics. The third possibility is to note that here, as in all particle-resolved work both experimental (colloids or granular media) and computational, the degree of supercooling is too limited to access the kind of growth in lengthcsales associated with a close approach to any transition ~\cite{royall2014physrep}. In particular the timescales we access approach the mode-coupling crossover and at deeper supercoolings different scaling behaviour may be encountered. Some evidence for the third possibility has recently been presented ~\cite{kob2011non,flenner2013}. Furthermore indirect measurements of dynamic correlation lengths obtained from a variety of experiments on molecular glassformers whose (relative) relaxation time is some \emph{ten decades} slower than particle-resolved studies access slow dynamic correlation lengths comparable to those we measure here ~\cite{berthier2005,dalleferrier2007,crauste2010,brun2011,bauer2013}. We thus hope that our work has gone some way to contributing to the debate on whether (local) structure may be related to dynamical arrest.

We have identified different forms of motion at similar state points. Such considerations have received some attention via the string-like motion~\cite{kob1997} and broken bond  ~\cite{kawasaki2013} concepts. Within the framework of Random First-Order Transition (RFOT) Theory~\cite{lubchenko2007}, one expects a crossover to more compact and less stringlike mobile regions at very deep supercooling~\cite{stevenson2006}. Consistent with this suggestion, a \emph{reduction} in certain measures of the dynamical correlation length at supercooling around the mode-coupling transition~\cite{kob2011non}. However later work with the more often used $\xi_4$ dynamic correlation length found no such reduction around the mode-coupling transition, rather a crossover to slower growth~\cite{flenner2013}. The ``two-dimensional'' motion we have identified here would correspond to a smaller dynamic correlation length for a given number of dynamically correlated particles. Tempting as it might be to draw analogies with the predictions of RFOT theory~\cite{stevenson2006}, we caution that our work only accesses relatively mild supercooling, up to around the mode-coupling crossover. The change in fractal dimension of dynamically fast regions envisaged by Stevenson \emph{et al.}~\cite{stevenson2006} corresponds to deeper supercooling than we access here.
Overall we believe our observation of different classes of motion discussed in section~\ref{sectionCorrelationOfDynamicsAndStructure} proposes an avenue that might be further investigated. We emphasise that not all motion takes the same form, here we have observed one and two-dimensional motion.

\subsection*{Acknowledgements} We acknowledge L. Cipelletti, D. Coslovich,  W. Kob, L. Ramos, H. Tanaka, A. Vailati and I. Williams for helpful discussions. CPR acknowledges the Royal Society 
and the European Research Council (ERC Consolidator Grant NANOPRS, project number 617266) for funding. \\


\end{document}